\documentclass{article}
\renewcommand{\epsilon}{\varepsilon}
\begin{document}
\title{Self-interacting gas in a gravitational wave field}
\author{Alexander B. Balakin \footnote{Electronic address:
Alexander.Balakin@ksu.ru}\\
Department of General Relativity and Gravitation\\
Kazan State University, 420008 Kazan, Russia\\
\ \\
Winfried Zimdahl\footnote{Electronic address:
winfried.zimdahl@uni-konstanz.de}\\
Fachbereich Physik, Universit\"at Konstanz\\
PF M678
D-78457  Konstanz, Germany}

\maketitle

\begin{abstract}
We investigate a relativistic self-interacting gas in the
field of an external {\it pp} gravitational wave.
Based on symmetry considerations we ask for those
forces which are able to compensate the imprint of the
gravitational wave on the macroscopic 4-acceleration of the gaseous fluid.
We establish an exactly solvable toy model according to which the
stationary states which characterize such a situation have negative entropy
production and are accompanied by instabilities of the microscopic particle motion.
These features are similar to those which one encounters in
phenomena  of self-organization in many-particle systems.
\end{abstract}
\ \\

\vspace{1.5cm}

\section{Introduction}
\label{introduction}

Let a gravitational wave (GW) act upon a simple relativistic gas
which was at ``global'' equilibrium (see,
e.g.,\cite{Stew,Ehl,IS,Groot}) before the infall of the GW. The
influence of the non-stationary radiation field on the gas
dynamics will generally destroy such an equilibrium. States of
``global'' equilibrium are only possible in stationary spacetimes,
i.e., in spacetimes which admit a timelike Killing vector (KV)
\cite{Stew,Ehl,IS,Groot,TauWei,Cher}. In the limiting case of
ultrarelativistic matter (radiation) the corresponding equilibrium
conditions imply the existence of a conformal timelike KV. Neither
a timelike KV nor a conformal timelike KV is compatible with the
properties of gravitational radiation. A {\it pp} GW is connected
with the symmetry groups $G _{3}$, $G _{5}$ and $G _{6}$, implying
a covariantly constant null vector field and a set of spacelike
KVs \cite{Petrov,Kraetal}. The gravitational radiation will
therefore induce non-equilibrium processes inside the gaseous
system which can be described with the help of linear irreversible
thermodynamics \cite{BaIg,BaGo,BaTro,BaAm}.

Typically, one studies the motion of particles or the behavior of continuous media under the influence of an infalling GW in order to reconstruct the properties of the latter.
This strategy relies on the circumstance that free particles, e.g., move on geodesics of the spacetime provided by the external field, and neighboring particles experience tidal forces which may be used to monitor the GW. Similar considerations hold for continuous media.

The universality of the gravitational interaction makes the behavior of a material system under the influence of an external gravitational field different from the corresponding behavior under the action of an external electromagnetic field.
In the latter case screening effects occur which may compensate the external action in the interior of the system.
There is no corresponding shielding of the gravitational interaction.

On this background one may, however, raise a different question: Is it possible to counterbalance (parts of) the imprint of an external GW by {\it non-gravitational} forces? Or, in other words, which kind of non-gravitational interaction might (at least partially)  ``shield'' the action of the gravitational field?

In the present paper we investigate this problem by admitting suitable self-interactions of a Maxwell-Boltzmann gas as candidates for such hypothetical screening.
The latter will be characterized by symmetry considerations as follows.
Before the GW infall the system is assumed to have certain spacetime symmetries, e.g. those of Minkowski space, which, in general, are broken by the GW. Our requirement for a (partial) compensation of the GW imprint on the gas will be that a part of the original symmetries continues to hold even under the GW influence.
Then we show, how this requirement can be realized with the help of the mentioned  interactions.

This strategy is comparable to the one used in gauge theories. Gauge field theories rely on the fact that local symmetry requirements neccessarily imply the existence of additional interaction fields, the gauge fields.
Although the symmetries in this paper are of a different type, the general feature, that symmetry requirements are accompanied by additional interactions, is true here as well.

On the other hand, there is no guarantee, that the required forces
really do exist. In case they do not exist, there is no shielding
of the described manner. But in any case, establishing the
self-interaction which would be necessary to realize a given
symmetry may improve our understanding of the dynamics of physical
systems in external gravitational fields.

This paper relies on a description of the gas by the relativistic Boltzmann equation
for the invariant one-particle distribution function.
According to the conventional picture the gas
particles follow geodesic worldlines of the radiative
spacetime in between elastic binary collisions.
Imposing symmetry conditions of the mentioned kind will, in general, make the particle trajectories differ from geodesics.
This difference to geodesic particle motion may be attributed to interactions which can be described either by an additional (in general inelastic) collision term or by an effective force on the individual particles.

The crucial point is that this interaction is not known in
advance. It depends on the external field and represents that
response which is necessary to realize the imposed symmetry. It is
equivalent to a self-interaction of the gas. Either there exist
corresponding interactions inside the gas or these forces are
mimicked by an external source. Both options are treated here on
an equal footing. If the system is unable to respond in the
required manner, the symmetry cannot be maintained. In other
words, the system has to organize itself properly in order to
counterbalance the external field successfully.

The considerations of this paper represent an exactly solvable toy model along these lines.
We determine the necessary interaction for a special case.
As an initial configuration we take a homogeneous and isotropic gaseous fluid in Minkowski space which macroscopically is at rest. Under these conditions it has a vanishing 4-acceleration $\dot{U}^i =0$.
In the field of a subsequently infalling GW the fluid motion will no longer be geodesic in general, i.e., one has $\dot{U}^i \neq 0$.
In the present case the mentioned symmetry requirement will be equivalent to {\it enforce a macroscopic
geodesic fluid motion $\dot{U}^i =0$ also under the influence of the GW field}.
This is achieved by allowing the trajectories of the individual {\it microscopic} gas particles to be {\it non-geodesic}.
(Without imposing the additional symmetry it would be vice versa, i.e., the microscopic trajectories were geodesic, but not the macroscopic fluid motion.)
The corresponding states are stationary, far-from-equilibrium states, accompanied by instabilities in the microscopic particle motion.
There exist striking similarities to processes of self-organization \cite{Prig,Hak,Klimont}.

The paper is organized as follows.
In Section \ref{force field} we determine the effective one-particle force on
microscopic gas particles which is compatible with an equilibrium distribution of the latter
(generalized equilibrium).
We establish our toy model by suitable symmetry considerations concerning the
vector $U^a /T$, where $U^a$ is the macroscopic 4-velocity and $T$ is the temperature.
In  Section \ref{transport equations}  we derive
the corresponding macroscopic balance equations for the first and second
moments of the one-particle distribution function
which, as a consequence of the force action, have nonvanishing sources in
general, characterizing the self-interaction of the system.
In Section \ref{particle motion} the equations of motion of the microscopic particles are
explicitly integrated.  The self-interaction
is shown to be accompanied by instabilities of the
transverse particle motion.
Our main results are summarized in Sec. \ref{summary}.
Units have been chosen so that $c = k _{B} = \hbar = 1$.

\section{Force field and generalized equilibrium}
\label{force field}

The one-particle distribution function $f = f\left(x,p\right)$ of a
relativistic gas is supposed to obey the Boltzmann type equation
\begin{equation}
L\left[f\right] \equiv
p^{i}f,_{i} - \Gamma^{i}_{kl}p^{k}p^{l}
\frac{\partial f}{\partial
p^{i}}
 = J\left[f,f\right] + {\cal S}\left(x, p\right)  \mbox{ , }
\label{1}
\end{equation}
where $J[f,f]$ is Boltzmann's collision integral. Anticipating that additional interactions will play a role in the following, we have included here a non-standard term ${\cal S}$ on the right-hand side, which takes into account those ``collisions'' which can not be reduced to elastic, binary interactions.
Following earlier applications in a cosmological context \cite{ZB1,ZB2,ZPRD98,ZSBP} we assume
${\cal S}$
to have the structure
\begin{equation}
{\cal S} = -m F ^{i}\frac{\partial{f}}{\partial{p ^{i}}}\ .
\label{2}
\end{equation}
It is straightforward to realize that a ``collision'' term of this form may be taken to the left-hand side of Boltzmann's equation (\ref{1}),  resulting in
\begin{equation}
p^{i}f,_{i} - \Gamma^{i}_{kl}p^{k}p^{l}
\frac{\partial f}{\partial
p^{i}} + m F ^{i}\frac{\partial{f}}{\partial{p ^{i}}}
 = J\left[f,f\right]   \mbox{ . }
\label{3}
\end{equation}
The left-hand side of this equation can be written as
\[
\frac{\mbox{d}f \left(x,p \right)}{\mbox{d}\gamma }
\equiv  \frac{\partial{f}}{\partial{x ^{i}}}
\frac{\mbox{d}x ^{i}}{\mbox{d}\gamma }
+ \frac{\partial{f}}{\partial{p ^{i}}}
\frac{\mbox{d}p ^{i}}{\mbox{d}\gamma}
\]
with
\begin{equation}
\frac{\mbox{d} x ^{i}}{\mbox{d} \gamma} = p ^{i}\ , \qquad
\frac{\mbox{D} p ^{i}}{\mbox{d} \gamma } = m F ^{i}\ .
\label{4}
\end{equation}
Equations (\ref{4}) are the equations of motion for gas particles which move under the influence of a force field $F ^{i}=F ^{i}\left(x,p \right)$.
The quantity $\gamma$ is a parameter along the particle worldline which for
massive particles may be related to the proper time $\tau $ by
$\gamma = \tau /m$.
Consequently, a specific ``collisional'' interaction, described by a ``source'' term ${\cal S}$, may be mapped onto an effective one-particle force $F ^{i}$.
This demonstrates that there exists a certain freedom to interpret collisional events in terms of forces. (This freedom can also been used in the reverse direction, i.e., to interprete (parts of forces) as collisions \cite{Kandrup}).
We emphasize that our approach is different from the ``canonical'' theory of particles in a force field for which the force term
$mF ^{i}\partial f/ \partial p ^{i}$  in Eq. (\ref{3}) is replaced by
$m\partial \left(F ^{i}f \right)/ \partial p ^{i}$ \cite{Kandrup}.
While both approaches are consistent with the equations of motion (\ref{4}), they coincide only for $\partial F ^{i}/ \partial p ^{i}=0$, which holds,
e.g., for the Lorentz force.
In the cases of interest here we will have
$\partial F ^{i}/ \partial p ^{i}\neq 0$.

For collisional equilibrium $f$ reduces to J\"uttner's distribution function
$f ^{0}$,
\begin{equation}
f \rightarrow f^{0}\left(x, p\right) =
\exp{\left[\alpha + \beta_{a}p^{a}\right] } \ ,
\label{5}
\end{equation}
where $\alpha = \alpha\left(x\right)$ and
$\beta_{a}\left(x \right)$ is timelike.
For $f \rightarrow f ^{0}$ the collision integral vanishes, i.e.,
$J \left[f ^{0},f ^{0}\right] = 0$.
Substituting $f ^{0}$ into Eq. (\ref{3}) we obtain
\begin{equation}
p^{a}\alpha_{,a} +
\beta_{\left(a;b\right)}p^{a}p^{b}
=  - m \beta _{i}F ^{i}
\mbox{ .}
\label{6}
\end{equation}
As already mentioned in the introduction, the situation here is
different than for example for charged particles in an
electromagnetic field where the microscopic starting point for a
self-consistent analysis is a given interaction, the Lorentz
force. Here, the analytic structure of the force is not known in
advance but has to be determined. This requires a kind of
``inverse'' strategy which is characterized by symmetry
considerations. In the present paper we are interested in that
specific force which realizes the requirement $U^i\pounds _{\beta}
g _{ik}  = 0$ where $\beta^i \equiv U^i /T$. This corresponds to
retaining a part of the original timelike symmetry $\pounds
_{\beta} g _{ik}  = 0$, by which the system was characterized
before the GW infall.

Once the analytic structure of the force has been determined on this basis,
the analysis may be carried out along the usual lines of a self-consistent
treatment.  Focusing on forces which satisfy the equilibrium condition
(\ref{6}), the most general structure of the term on the right-hand side of
equation (\ref{6}) is
\begin{equation}
\beta _{i}F ^{i} =
\beta _{i}F ^{i}_{a}\left(x \right)p ^{a}
+ \beta _{i}F ^{i}_{ab}\left(x \right)p ^{a}p ^{b} \ .
\label{7}
\end{equation}
Comparing now different powers in Eq. (\ref{6}) separately, we obtain the
(generalized) equilibrium conditions
\begin{equation}
\alpha _{,a} = - m \beta _{i}F _{a}^{i}
\label{8}
\end{equation}
and
\begin{equation}
\beta _{\left(k;l \right)}
\equiv  \frac{1}{2}\pounds _{\beta}g _{kl} = - m \beta _{i}F^{i}_{kl}\
.
\label{9}
\end{equation}
The last equation relates space-time symmetries to the force exerted on the
particle.
The Lie derivative may be invariantly split into contributions
parallel and perpendicular to the fluid four-velocity:
\begin{equation}
\pounds _{\beta} g _{ik}
= - U_i U^n\pounds _{\beta} g _{nk}
+ h_i ^n \pounds _{\beta} g _{nk}
\ ,
\label{10}
\end{equation}
where $h _{ik} \equiv  g _{ik} + U_{i}U_{k}$,
\begin{equation}
U^n\pounds _{\beta} g _{nk}
= - 2\frac{\dot{T}}{T ^{2}}U_k
+ \frac{1}{T}\left[\dot{U}_{k} + \frac{\nabla _{k}T}{T} \right]
\ ,
\label{11}
\end{equation}
and
\begin{equation}
h_i^n\pounds _{\beta} g _{nk}
= - \frac{U_k}{T}\left[\dot{U}_{i} + \frac{\nabla _{i}T}{T} \right]
+ \frac{2 \sigma _{ik}}{T}
+ \frac{2}{3}\frac{\Theta}{T} h _{ik}
\ .
\label{12}
\end{equation}
Here, $\dot{T} \equiv  T _{,a}U^{a}$ is the derivative in direction of
the four-velocity, $\dot{U}_{a} \equiv  U_{a;c}U^{c}$ is the
four-acceleration,
$\nabla  _{a}T \equiv  h _{a}^{b}T _{,b}$ is the covariantly defined spatial
derivative,
$\Theta \equiv  U^{i}_{;i}$ is the fluid expansion scalar, and
$\sigma _{ab}$ is the shear tensor.
The quantity $U^{a}/T$ is a Killing-vector (KV) if both parts (\ref{11})
and (\ref{12}) in (\ref{10})
are zero.
Let us assume that the initially homogeneous and isotropic gas
is at rest and in equilibrium
and that $\pounds  _{\beta}g _{ik}  = 0$ holds
before the GW infall.
Generally, this symmetry will be destroyed by the GW.
The action of a GW on matter is known to produce anisotropies and
inhomogeneities within a test medium of this kind.

Now we ask, whether and under which conditions it is possible
to retain a part of the latter symmetry, namely the projection
\begin{equation}
u^i\pounds _{\beta} g _{ik}  = 0
\ ,
\label{13}
\end{equation}
also in the GW field.
According to (\ref{11}) this condition implies
$\dot{T}=0$ and
$\dot{U}_{m} + \frac{\nabla  _{m}T}{T}=0$.
This is equivalent to retaining that part of the previous equilibrium
conditions which fixes the temperature behavior but to allow the quantities
$\Theta$ and $\sigma _{ab}$ to be different from zero.
This case turns out to be exactly solvable as will be shown in the remainder
of the paper.
For the special case $ \nabla _{m}T =0$ the above condition implies
$\dot{U}_{m}=0$, i.e., a  geodesic fluid motion.

Imposing the corresponding ``partial symmetry'' condition
\begin{eqnarray}
u^i\pounds _{\beta} g _{ik}  = 0 \Leftrightarrow \pounds _{\beta} g _{ik}  &=&
\frac{{\rm 2}}{{\rm 3}}\frac{{\rm \Theta }}{T} h _{ik}
+ \frac{{\rm 2}}{T}\sigma  _{ik}\nonumber\\
&=& \frac{{\rm 1}}{T}\left[\nabla _{k}U_{i} + \nabla  _{i}U_{k}\right]
\equiv  \frac{{\rm 2\Theta } _{ik}}{T}\ ,
\label{14}
\end{eqnarray}
is consistent with a non-vanishing quantity $U_{i}F ^{i}$ in which
\begin{equation}
mU_{i}F ^{i}_{ab} = - \Theta _{ab}\ ,
\label{15}
\end{equation}
i.e., this condition fixes the part which is quadratic in the
particle momenta.
Obviously, macroscopic fluid quantities enter the analytic expression for the
force which is exerted on the microscopic particles.
The effective interaction couples the microscopic particle
momenta with quantities such as the fluid shear tensor which characterizes
the system on a macroscopic level.
This is quite  similar to the well known self-consistent coupling of the
particle momentum of a charged particle to the electromagnetic field
strength
tensor.
The linear part of the force is determined by the derivative of the quantity $\alpha$
according to
\begin{equation}
- \frac{m}{T} U_{i}F ^{i}_{a} = - \dot{\alpha}U_{a} + \nabla  _{a} \alpha \
.
\label{16}
\end{equation}
The partial symmetry condition (\ref{14}) and Eq.~(\ref{16}) constitute the conditions of
``generalized equilibrium'' for
the system of gas particles in the force field (\ref{7}).
In the force-free limit these conditions coincide with those for
``global'' equilibrium \cite{Stew,Ehl,Groot}.

An example for a force linear in $p ^{a}$ is the Lorentz force which is
obtained for $F ^{ab}\rightarrow \frac{e}{m}F ^{ab}_{\left(em \right)}$
where $e$ is the charge and
$F ^{ab}_{\left(em \right)}$ is the electromagnetic field strength tensor.
For this case the equilibrium condition (\ref{8})  is well known in the
literature
\cite{Ehl,Groot}. Other cases have been discussed in a cosmological context
\cite{ZB1,ZB2,ZPRD98,ZSBP}.

The conditions (\ref{15}) and (\ref{16})  do not yet fix the spatial
part of the force.
The condition for the latter is that the force itself has to be orthogonal
to the four momentum, i.e., $p_iF ^{i}= 0$ .
This requirement is satisfied for the choice
\begin{equation}
mF ^{i}_{ab} p ^{a}p ^{b}= \left[\Theta _{ab}U^{i}
- \Theta ^{i}_{a}U_{b}\right]p ^{a}p ^{b}\ .
\label{17}
\end{equation}
For particles moving exactly with the macroscopic fluid velocity, i.e.,
$p ^{k}=mU^{k}$, this force vanishes.
It follows that the force is related to the relative motion of the
individual particle and the fluid as a whole.
In a sense this is similar to the motion of a sphere in a viscous fluid
described by Stokes' formula.
The spatial part of this contribution to the force is
\begin{equation}
mh _{i}^{n}F ^{i}_{ab} p ^{a}p ^{b} = \Theta ^{n}_{a}Ep ^{a}
\ ,
\label{18}
\end{equation}
the quantity $E \equiv  -U_{a}p ^{a}$ being the energy of the particle for a comoving
observer.
In the nonrelativistic limit with $E \approx m$ we obtain
\begin{equation}
h _{i \nu }F ^{i}_{ab} p ^{a}p ^{b}
\stackrel{nr}{\longrightarrow }
\frac{1}{2}
\left(v _{\nu ,\mu } + v _{\mu ,\nu }\right)
p ^{\mu }
\ ,
\label{19}
\end{equation}
where $v _{\mu }$ is the nonrelativistic fluid velocity and $\mu ,
\nu = 1, 2, 3$. This force couples the particle velocity $p ^{\mu
}/m$ to the velocity derivative of the medium through which the
particle moves. It represents a kind of friction. In this analogy
the symmetrized spatial derivative of the fluid velocity plays the
role of a friction ``constant''. The difference to the motion of a
nonrelativistic particle in a viscous fluid is the self-consistent
nature of the problem in our case: The particle which is subject
to the force exerted by the medium is part of this medium itself.
In the limit of Eq. (\ref{19})  this force represents a linear
coupling of the particle momentum and the derivatives of the fluid
velocity. In a sense, this parallels the Lorentz force which
couples the particle momentum linearly to the derivatives of the
electromagnetic potentials.

Finishing this section we emphasize again the
unconventional strategy which was used here to obtain the
analytical properties of the force $F ^{i}$: The latter quantity
which is similar to a friction force characterizes just those interactions which
guarantee the prescribed (partial) symmetry condition
(\ref{14}) of our model.

\section{Transport equations}
\label{transport equations}

The particle number flow 4-vector
$N^{i}$ and the energy momentum tensor $T^{ik}$ are
defined in a standard way (see, e.g., \cite{Ehl,IS}) as
\begin{equation}
N^{i} = \int \mbox{d}Pp^{i}f\left(x,p\right) \mbox{ , }
\ \
T^{ik} = \int \mbox{d}P p^{i}p^{k}f\left(x,p\right) \mbox{ .}
\label{20}
\end{equation}
The integrals in the definitions (\ref{20}) and in the following
are integrals over the entire mass shell
$p^{i}p_{i} = - m^{2}$.
The entropy flow vector $S^{a}$ is given by \cite{Ehl,IS}
\begin{equation}
S^{a} = - \int p^{a}\left[
f\ln f - f\right]\mbox{d}P \mbox{ , }
\label{21}
\end{equation}
where we have restricted ourselves to the case of
classical Maxwell-Boltzmann particles.
The entropy production density is a sum of two terms:
\begin{equation}
S ^{i}_{;i}  = \sigma _{c} + \sigma _{F}\ .
\label{22}
\end{equation}
Here,
\begin{equation}
\sigma_{c} \equiv - \int \mbox{d}P
J\left[f,f\right]  \ln f
\label{23}
\end{equation}
is the familiar contribution of Boltzmann's collision integral, while
\begin{equation}
\sigma_{F} \equiv    m\int \mbox{d} P
F ^{i}\frac{\partial{f}}{\partial{p ^{i}}}\ln f
\label{24}
\end{equation}
takes into account an entropy production due to the action of the force
$F ^{i}$.
Since Boltzmann's $H$ theorem guarantees $\sigma _{c} \geq 0$,
we have
\begin{equation}
S^i _{;i} - \sigma _{F} \geq 0 \ .
\label{25}
\end{equation}
The equality sign in the last relation is realized for  $f \rightarrow f ^{0}$.
With $f$ replaced by $f^{0}$ in the definitions
(\ref{20}) and (\ref{21}), the quantities $N^{a}$, $T^{ab}$ and $S^{a}$ may be
split with respect to the unique 4-velocity $U^{a}$ according to
\begin{equation}
N^{a} = nU^{a} \mbox{ , \ \ }
T^{ab} = \rho U^{a}U^{b} + p h^{ab} \mbox{ . \ \ }
S^{a} = nsU^{a} \mbox{  .}
\label{26}
\end{equation}
The balances
\begin{eqnarray}
N^{a}_{;a}&=&-m \beta _{i}\int F ^{i}f ^{0}\mbox{d}P \ , \nonumber\\
T^{ak}_{\ ;k}&=&-m \beta _{i}\int p^{a}F ^{i}f ^{0}\mbox{d}P \ ,
\label{27}
\end{eqnarray}
may be written as
\begin{equation}
N ^{a}_{;a} = - m \beta _{i}\left(F ^{i}_{a}N ^{a}
+ F ^{i}_{ab}T ^{ab} \right)\ ,
\label{28}
\end{equation}
and
\begin{equation}
T ^{ak}_{\ ;k} = - m \beta _{i}\left(F ^{i}_{b}T ^{ab}
+ F ^{i}_{kl}M^{akl} \right)\ ,
\label{29}
\end{equation}
where $M ^{akl} = \int \mbox{d}P f ^{0}p ^{a}p ^{k}p ^{l}$ is the third
moment of the equilibrium distribution function.
For the entropy production density we find
\begin{equation}
S^{a}_{;a} = m \beta _{i} \int \left[\alpha + \beta _{a}p ^{a} \right]
F ^{i}f ^{0}
\mbox{d}P
=  - \alpha N^{a}_{;a}
- \beta_{a}T^{ab}_{\ ;b}
\mbox{ . }
\label{30}
\end{equation}
Evaluating the integrals in the balances (\ref{28}) and (\ref{29})  we
obtain
\begin{equation}
N ^{a}_{;a} = n \dot{\alpha} + n \Theta \equiv  n \Gamma \ ,
\label{31}
\end{equation}
\begin{equation}
U_{a}T ^{ak}_{\  ;k} = - \dot{\alpha}\rho
- \left(\rho + p \right)\Theta \ ,
\label{32}
\end{equation}
and
\begin{equation}
h _{ca}T ^{ak}_{\  ;k} = p\nabla  _{c}\alpha \ .
\label{33}
\end{equation}
Before the GW infall the gas was supposed to be at rest and the initial
timelike symmetry implied
$\dot{n}=\dot{T}=\dot{\rho }=\dot{s}=0$.
Now we ask under which conditions the quantities
$n$, $T$, $\rho$ and $s$ may be constant in time also under
the GW influence.
Obviously, this requires a force which guarantees
$\dot{\alpha}=0$.
Indeed, as to be seen from Eqs. (\ref{31}) and (\ref{32})  with (\ref{26}),
the latter condition implies $\dot{n}=\dot{\rho }=0$, while
$\dot{T}=0$ is already part of our basic assumptions
[cf. the discussion preceding Eq. (\ref{14})].
From the Gibbs equation
\begin{equation}
T \mbox{d}s = \mbox{d} \frac{\rho }{n} + p \mbox{d}\frac{1}{n}
\quad\Rightarrow\quad n T \dot{s} = \dot{\rho } - \left(\rho + p \right)
\frac{\dot{n}}{n}
\label{34}
\end{equation}
it follows, that $\dot{s}=0$ is valid as well.
However, the force which maintains the
equilibrium distribution gives rise to source terms in the balances
(\ref{31}) -- (\ref{33}).
Two interpretations are possible here. Either the sources (right-hand sides) in (\ref{32}) and (\ref{33})
may be taken to the left-hand sides of these equations and regarded as stress components
of a conserved energy-momentum tensor for a self-interacting medium [cf. \cite{ZB1,ZB2,ZPRD98,ZSBP}],
or they describe an equivalent, but externally driven interaction
which effectively is a self-interaction as well.
Dynamically, both interpretations are indistinguishable. The interaction terms
describe the support from internal or external sources which is necessary
to resist the  ``attack'' of the GW.
However, our toy model leaves open the question, how these interactions my be realized.
For the part $\sigma _{F}$ of the entropy production [cf. Eq. (\ref{24})] we
find  under these circumstances
\begin{equation}
\sigma _{F} = ns \Theta \ ,
\label{35}
\end{equation}
where we have used the expression
\begin{equation}
s = \frac{\rho + p}{nT} - \alpha \
\label{36}
\end{equation}
for the entropy per particle together with
the identification $\alpha = \mu /T$ where $\mu $  is the chemical
potential. We emphasize that our generalized equilibrium due
to the partial symmetry requirement (\ref{14})
is consistent with a nonvanishing entropy production density.
While the contribution $\sigma _{c}$ [cf. Eq. (\ref{23})] to the total entropy production density (\ref{22}) vanishes for
$f \rightarrow f ^{0}$, the contribution $\sigma _{F}$  [cf. Eqs. (\ref{24}) and (\ref{35})] does not.
The corresponding states are out-of-equilibrium configurations of a specific kind.
Generally, the theoretical description of out-of-equilibrium phenomena is known to be a major problem as soon as the deviations from a standard equilibrium reference state cannot be considered as small.
The concept of generalized equilibrium is an approach to characterize a certain class of states which may be far away from ``conventional'' equilibrium (vanishing entropy production), but nevertheless are described by a J\"uttner function and, consequently, admit an exact analytic treatment.
Macroscopically, the imposed (partial) symmetry implies
\begin{equation}
S^k _{;k} = n s \Theta \ ,
\mbox{\ \ \ \ }
\dot{n} = \dot{\rho } = \dot{s} = 0 \ .
\label{37}
\end{equation}
This means that the equilibrium properties (\ref{5}) and $\dot{n} = \dot{T}=\dot{\rho } = \dot{s} = 0 $, which are supposed to characterize the system before the GW infall, remain valid also under the influence of the gravitational radiation.
It is obvious from Eq. \ref{31}) that for such a situation to be realized, a change in the number of particles
is necessary.
The existence of source terms in the balances for the moments of the distribution function
suggests an interpretation of the gas as an ``open'' thermodynamical system.
It is well known that stationary states of open systems away from ``conventional'' equilibrium are key ingredients for phenomena of self-organization (see, e.g., \cite{Prig,Hak,Klimont} and references therein).

\section{Particle motion}
\label{particle motion}

\subsection{Energy conservation and expansion scalar}

In this section we study the microscopic particle motion in the force field
(\ref{17}) under the influence of an external gravitational wave.
According to Eqs. (\ref{4}), (\ref{7}), (\ref{17}) and neglecting the linear part of the force, i.e., $\beta_iF^i_a = 0$ we have to solve the system
\begin{equation}
\frac{D p_i}{D \tau} =
\frac{1}{m} \left[\Theta_{kl} U_i - \Theta_{il}  U_k \right] p^k p^l  \ ,
\mbox{\ }
\frac{dx^k}{d\tau} = \frac{p^k}{m}\
\label{38}
\end{equation}
in the background metric
\begin{equation}
d\tau^2 = - 2dudv  + g_{\alpha \beta}(u) dx^{\alpha}dx^{\beta} \ ,
\label{39}
\end{equation}
where
\begin{equation}
\alpha, \beta = 2,3\ ,
\mbox{\ \ \ \ }
u = \frac{t - x^1}{\sqrt{2}}\ ,
\mbox{\ \ \ \ } v = \frac{t + x^1}{\sqrt{2}}\ .
\label{40}
\end{equation}
This metric is an exact solution of Einstein's vacuum equations
with a $G_5$ symmetry group (see, e.g., \cite{Petrov}, \S 33,
\cite{Kraetal}, chapter 21.5).
It admits five Killing
vector, three of which, namely,
\begin{equation}
\xi^i_{(v)} = \delta^i_v \ , \ \ \
\xi^i_{(2)} = \delta^i_2\ , \ \ \
\xi^i_{(3)} = \delta^i_3\ ,
\label{41}
\end{equation}
being the generators of an Abelian subgroup $G_3$.
The first of the vectors (\ref{41}) is covariantly constant and isotropic,
$\xi^i_{(2)}$ and $\xi^i_{(3)}$ are spacelike.
Evidently, the dynamic equations (\ref{38}) admit the  quadratic integral
of motion $p_i p^i = - m^2 $.
To find another integral of motion let us realize that
\begin{equation}
\frac{D}{D\tau}(U^i p_i)  = p^i \frac{p^l}{m} U_{i;l} +
U^i F_i = - \frac{1}{m} p^k p^l  U_k \dot{U}_l \ .
\label{42}
\end{equation}
In order to make further analytical progress, we first realize
that the KVs (\ref{41}) of the radiative spacetime are also part
of the symmetries of the initial configuration before the GW
infall. Under this condition it is tempting to assume that the
fluid quantities ``inherit'' the spacetime symmetries
\cite{BaAm,BaPo}. In particular, we require that the symmetries
$\pounds _{\xi_{(r)}} g _{ik} = {\rm 0}$ for $r = v,2,3$,
corresponding to the Killing vectors (\ref{41}), are inherited by
the temperature, i.e., $\pounds_{\xi_{(r)}}T = {\rm 0}$. Combining
this kind of temperature behavior along the vectors (\ref{41})
with the equilibrium condition $\dot{T} = 0$  implies a constant
temperature both in space and time. According to the condition
$\dot{U}_{m} + \frac{\nabla  _{m}T}{T}=0$, however, a spatially
constant temperature is equivalent to $\dot{U}_{i} = 0$.
Consequently, a combination of the  condition  $U^i\pounds
_{\beta}g _{ik}  = 0$ with the requirement $\pounds_{\xi_{(r)}}T =
{\rm 0}$ of symmetry inheritance implies a geodesic fluid motion.
Under these circumstances we have
\begin{equation}
\frac{\mbox{D} E}{\mbox{D}\tau}  = 0 \ ,
\mbox{\ \ \ \ \ }
E \equiv - U_i p^i = {\rm const} \ ,
\label{43}
\end{equation}
i.e., the quantity $E$, which may be interpreted as particle
energy for a comoving (with $U ^{i}$) observer, is a second
integral of motion. Contracting $\dot{U}_{i}$ with the Killing
vectors (\ref{41}) leads the following well-known consequences
\cite{Ehl}:
\begin{eqnarray}
\xi^i_{(r)} \dot{U}_i     &=& U^k (\xi^i_{(r)}U_i)_{;k}   -
\frac{1}{2}U^i U^k \left( \xi_{k;i} + \xi_{i;k} \right) \nonumber\\
&=&
(\xi^i_{(r)}U_i)^{\displaystyle \cdot} = 0 \ ,
\label{44}
\end{eqnarray}
equivalent to
\begin{equation}
\xi^i_{(r)}U_i = {\rm const}\ ,
\mbox{\ }  U_r = U_r \left(0 \right)\ .
\label{45}
\end{equation}
It follows that the unit time-like velocity vector in
the GW field has the form
\begin{equation}
U_i = \delta_i^v U_v(0)+ \delta_i^2 U_2(0) + \delta_i^3 U_3(0)
+ \delta_i^u  U_u(u)\ ,
\label{46}
\end{equation}
where
\begin{equation}
U_u(u) =  \frac{1 + g^{\alpha\beta}(u)  U_{\alpha}(0)
U_{\beta}(0)}{2 U_v(0)}\ .
\label{47}
\end{equation}
This solution, which satisfies the symmetry inheritance condition
$\pounds_{\xi_{(r)}} U_i  = {\rm 0} $ for $r = v,2,3$,  allows us to
calculate the expansion scalar $\Theta $:
\begin{equation}
\Theta  =  \frac{1}{\sqrt{-g}}
\frac{\mbox{d}}{\mbox{d}u}
\left(U^u \sqrt{-g} \right) = -U_v(0) \frac{\mbox{d}}{\mbox{d}u} \ln
\sqrt{-g}\ .
\label{48}
\end{equation}
Here, $U^u=-U_v$ is a non-negative  quantity  by definition and  $-g$, where $g$ is the
determinant of the metric tensor, is monotonically decreasing with $u$ (see,
e.g., \cite{MTW}, \S 35.9).
We find that $\Theta < 0$, i.e., the expansion scalar is negative, which
makes the present situation completely different from previous studies of
generalized equilibrium in a cosmological context \cite{ZB1,ZB2,ZPRD98,ZSBP}.
According to the balance equation (\ref{31}) (with $\dot{\alpha}=0$), a
negative expansion scalar implies the
{\it annihilation} of particles.
As a consequence, we have negative entropy production as well
[cf. Eq. (\ref{37})].
We emphasize that this kind of behavior does not contradict the second law
of thermodynamics since the gas under consideration here is an open
system (see the discussion following Eq. (\ref{37})).
Loss of entropy in an open system is typical for processes of
self-organization \cite{Prig}.
We argue that the mobilization of suitable forces to maintain the
relations
(\ref{5}) and (\ref{37}) in the presence of the GW exhibits features
of a self-organizing system.
The situation may be illustrated by a martial picture:
Under the ``siege'' of the GW the system will ``evacuate'' part of its
population in order to maintain partially its previous internal state.
Unless the GW stops to act upon the system this process continues until the
system has lost all its particles.
On the other hand,
if the system is unable to organize suitable supporting forces, it will not
succeed in further following its own internal dynamics but it will be
determined by the external field.
If there are no forces mobilized at all, the system is entirely ``slaved''
and its particles have to move entirely as prescribed by the external agent,
i.e., along the geodesics of the radiative spacetime.

\subsection{Longitudinal motion}

By using relations (\ref{43}) and (\ref{47})
the components $p_u = - p^v$ and $p_v = - p^u$
may be expressed
in terms of $p_{\alpha}$ and the components of
$U _{i}$:
\begin{equation}
p_v =  \frac{1}{2U_u} \left[\epsilon
\pm \sqrt{\epsilon^2
- \left(1+ g^{\alpha\beta} U_{\alpha}U_{\beta} \right)
\left(m^2 + g^{\alpha\beta} p_{\alpha}p_{\beta} \right)} \right]
\label{49}
\end{equation}
with the abbreviation
\[
\epsilon \equiv  E - g^{\alpha\beta} U_{\alpha}p_{\beta}
\]
and relation
\begin{equation}
2p_u p _{v}= m^2 + g^{\alpha\beta}  p_{\alpha}p_{\beta}\ .
\label{50}
\end{equation}
With the well-known  interpretation   of   the   covariantly
constant  isotropic  Killing  vector $\xi^i_{(v)}$ as
wave vector of  the  gravitational wave (see, e.g., \cite{Zakh}, chapter 9,
\cite{Steph}, chapter 15.3),
the projection $p_v = p_i \xi^i_{(v)}$ of the particle
momentum onto the GW wave vector represents the
``longitudinal'' component with respect to  GW  null  direction.
This scalar quantity will play
an important role in the further integration process.
The signs minus or plus in the expression (\ref{49}) correspond to the cases
where
the particle momentum has positive or negative projection, respectively,
onto the spatial propagation direction  of  the  gravitational  wave.
This may be most simply demonstrated if the system as a whole is at rest,
i.e.,
\begin{equation}
U_2 = U_3 = 0, \ U_1 \equiv \frac{U_v - U_u}{\sqrt{2}}
= 0\ , \ U_u = U_v = - \frac{1}{\sqrt{2}}\ ,
\label{51}
\end{equation}
and consists of massless particles, moving parallel or anti-parallel to the GW propagation
direction ($x ^{1}$-direction) with $p _{\alpha}=0$.
Under these circumstances we obtain
\begin{equation}
p_v = - p^{u} = \frac{- p^0 + p^1}{\sqrt{2}} =
\frac{- \vert  p^1 \vert + p^1}{\sqrt{2}}  =
\frac{- E \mp E}{\sqrt{2}}\ .
\label{52}
\end{equation}
The projection  $p_v$ vanishes for a positive  $p ^{1}$, which corresponds
to the minus sign in formula (\ref{49}).
The plus sign in formula (\ref{49}) together with the fact that
$U _{u}$ is negative according to Eq. (\ref{51}), corresponds to a negative
projection $p _{v}$, consistent with Eq. (\ref{52}).

\subsection{Transversal motion}

According to Eqs. (\ref{38}) and (\ref{43}) for $U_\alpha \equiv 0$, the
components $p_{\alpha}$ satisfy the nonlinear equations
\begin{equation}
\frac{\mbox{d} p_{\alpha}}{\mbox{d}\tau} =
\frac{E}{m} \Theta_{\alpha k} p^k \ .
\label{53}
\end{equation}
Their evolution is  coupled to  $p_v$ and  $p_u$, which are given by
formulas
(\ref{49}) and (\ref{50}), respectively.
In order to simplify the analysis, we assume that the gas was
macroscopically at rest before the infall of the GW.
The propagation direction of the GW will be the only preferred direction in
such a case and, we will refer to the components of the particle momentum
perpendicular to this direction as ``transversal''.
If so, the relevant components of the tensor $\Theta _{ik}$ are
\begin{equation}
\Theta_{\alpha u} = \Theta_{\alpha v} = 0\ ,
\mbox{\ \ \ \ \ }
\Theta_{\alpha \beta}   =    \frac{1}{2\sqrt{2}}
g'_{\alpha \beta}(u)\ ,
\label{54}
\end{equation}
where the prime denotes the derivative with respect to $u$.
Substituting the  derivatives with respect to affine parameter
$\tau$ by
\begin{equation}
\frac{\mbox{d}}{\mbox{d}\tau} = \frac{\mbox{d}u}{\mbox{d} \tau}
\frac{\mbox{d}}{\mbox{d}u} =
\frac{p^u}{m}\frac{\mbox{d}}{\mbox{d}u} =
- \frac{p_v}{m}\frac{\mbox{d}}{\mbox{d}u}\ ,
\label{55}
\end{equation}
we obtain from Eqs. (\ref{49}), (\ref{50}), and (\ref{53}) the  following
two-dimensional nonlinear, nonautonomous dynamic system:
\begin{eqnarray}
\frac{\mbox{d} p_{\alpha}}{\mbox{d}u} &=& \frac{1}{2}
\left[g'_{\alpha \beta}(u) g^{\beta \gamma}(u) \right]\cdot \nonumber\\
&&\ \ \cdot\frac{E p_{\gamma}}
{ E   \pm  \sqrt{E^2  - m^2 - g^{\alpha \beta}(u) p_{\alpha}
p_{\beta} }}\ .
\label{56}
\end{eqnarray}
A special solution of the last equation is
\begin{eqnarray}
&&p_v  = - \frac{E  \pm \sqrt{E^2 - m^2 }}{\sqrt{2}}\ , \
p_u  =  - \frac{E  \mp \sqrt{E^2 - m^2}}{\sqrt{2}}\ , \nonumber\\
&&p_{\alpha} =  0\ , \
p_1  = \mp  \sqrt{E^2 - m^2}\ .
\label{57}
\end{eqnarray}
All components of the momentum are constant in this case, which corresponds to a vanishing
of the self-interacting force on
the right-hand side of Eq. (\ref{53}) with the tensor components (\ref{54}).
The solution (\ref{57}) is consistent with
Eqs. (\ref{49}), (\ref{50}) and (\ref{52}).

\subsubsection{Linear stability analysis}

In a next step we consider the stability of the solution (\ref{57}) in the
vicinity of $p_{\alpha}  =  0$.
To this purpose we linearize the dynamic system (\ref{56}) with
respect to small deviations from zero ($\vert  p_{\alpha}\vert
<< m$). For simplicity we assume that $g_{23} =
0$.
Neglecting quadratic terms in $p_{\alpha}$
we find, that the expressions (\ref{57}) for  $p_u$ and $p_v$ remain valid
up to first order in the deviations from $p_{\alpha}  =  0$.
The corresponding solutions are
\begin{equation}
p_2(u) = C_2 \vert g_{22}(u) \vert^{\Gamma(E)}\ ,
\mbox{\ }
p_3(u) = C_3 \vert g_{33}(u) \vert^{\Gamma(E)}\ ,
\label{58}
\end{equation}
with
\begin{equation}
\Gamma(E) \equiv \frac{E
\left(E \mp \sqrt{E^2 - m^2} \right)}{2 m^2}\ .
\label{59}
\end{equation}
Obviously, $\Gamma(E)$ is  positive definite.
Recalling that the lower sign in the last expression corresponds to a
particle motion parallel to the spatial propagation direction of the GW,
$\Gamma(E)$ increases
monotonically  from $\frac{1}{2}$ for $E=m$, to
$\left(\frac{E}{m}\right)^2$, for  $E >> m$.
If the particle moves in the opposite direction,
$\Gamma(E)$
decreases monotonically  from $\frac{1}{2}$ for $E=m$, to
$\frac{1}{4}$ for $E >> m$.
The stability  properties depend on the values of $\vert g_{22}(u) \vert$
and
$\vert g_{33}(u) \vert$.
In the absence of the GW one has
$\vert  g_{22}(u)  \vert = \vert g_{33}(u) \vert = 1$ with
$p_2(u)  =  C_2$ and $p_3(u)  =  C_3$,  independently of
$\Gamma(E)$.
In  the GW field, one of the functions $\vert
g_{22}(u) \vert$ or $\vert g_{33}(u) \vert$ is larger, the other one smaller
than $1$.
For example, let us consider the well-known exact solution of
Einstein's  equations, where  $g_{22}=  \cos^2ku$ and $g_{33}= \cosh^2ku$.
(This solution is obtained from formula (25.23) in
\cite{Petrov} by substituting
$ \bar{c}=\bar{\bar{c}} =0$,  $\alpha = -1$, $x ^{4} = ku$ and
$x ^{1} = v/k$, and by changing the metric signature.)
The perturbations of the component $p _{2}$ will decrease in this case,
while there is a dynamical instability in the $p _{3}$ component.
The corresponding growth rate depends on the energy $E$
and on the direction of the particle motion  with  respect
to the spatial propagation direction of the GW.
We conclude that the self-interaction which is
necessary to maintain the (generalized) equilibrium under the
influence of the external GW, is connected with an instability of
the transverse particle motion. Dynamical instabilities are known
to accompany processes of self-organization (see, e.g.,
\cite{Prig,Hak,Klimont}). Although a more detailed correspondence to
such kind of effects in synergetics has still to be worked out,
the existence of instabilities in the particle motion seems to
support again the interpretation that self-interaction in the
present context exhibits features  of
self-organization.

\subsubsection{Exact one-dimensional solution}

With the above result of the linear perturbation analysis that the particle
motion in $x^3$ direction  is  unstable, while there is stability
along  the  $x^2$ axis, it seems reasonable to investigate a
model with $p_2 \equiv 0$, but $p_3 \neq 0$.
The equation to be solved in this case is
\begin{equation}
\frac{\mbox{d}p_3}{\mbox{d}u} = p_3 \frac{E}{2} \frac{(g_{33})'}{g_{33}}
\left[E \pm  \sqrt{E^2 - m^2 - \frac{(p_3)^2}{g_{33}}} \right]^{-1} \ .
\label{60}
\end{equation}
Using again the solution $g_{33} = \cosh ^{2}{ku}$ and substituting
\begin{equation}
\vert p_3 \vert = \cosh{ku} \sqrt{E^2 - m^2 - \left(Z - E \right)^2}\ ,
\label{61}
\end{equation}
the function $Z$ is implicitly given by the
solution of the transcendent equation
\begin{equation}
\left[\lambda  - Z + E \right]^{\frac{E}{\lambda }+1}
\left[\lambda  + Z - E \right]^{- \frac{E}{\lambda }+1} =
\frac{{\rm const ^{2}}}{\cosh^2ku}\ ,
\label{62}
\end{equation}
where $\lambda ^2$  denotes the abbreviation
$\lambda ^2 \equiv E^2 - m^2$ .
An exact analysis is possible for massless  ($m=0$) or ultrarelativistic ($m
\to 0$) particles. In this case one obtains
\begin{equation}
Z = 2 E - \frac{{\rm const}}{\cosh{ku}}\ ,
\label{63}
\end{equation}
and
\begin{equation}
p_3^2 = C_3^2 + 2 E (\cosh{ku} - 1) \left(E \pm
\sqrt{E^2 - C_3^2} \right)\ .
\label{64}
\end{equation}
The integration constant has been chosen such that $p_3 =
C_3$ at $u=0$.
The trivial solution $p _{3} \equiv  0$ of Eq. (\ref{60}) is realized in the
expression (\ref{64}) for
$C_3=0$ with the minus sign in front of the square root on the right-hand
side of this equation.
The exact, explicit solution (\ref{64}) shows that $p_3$ really increases
with
the  retarded  time  $u$.
It confirms the result of the previous linear analysis according to which
the transverse particle motion is unstable.
It should be mentioned, however, that the  quantity
$\frac{p_3^2}{\cosh^2ku}$,  decreases, so that the
expression under the square root on the right-hand-side of formula
(\ref{60}) remains positive.

\section{Summary}
\label{summary}

Given that a homogeneous, isotropic fluid at rest in Minkowski
space admits a timelike Killing vector $U^a /T$, we studied the
behavior of this system under the condition that the projection
$U^i\pounds_{\beta}g _{ik}$ of the Lie derivative of the metric in
direction of $U^i$ remains zero also in the GW field, i.e.,
$U^i\pounds_{\beta}g _{ik}  = 0$. This restricted timelike
symmetry requires the existence of interactions which we have
modelled as a self-interacting effective one-particle forces that
is exerted on the microscopic particles of a Maxwell-Boltzmann
gas. As a toy model we investigated the exactly integrable special
case in which these forces result in compensating the GW imprint
on the macroscopic fluid acceleration. We suggested an
interpretation of the gas as an open thermodynamical system. In
particular, the considered interaction is accompanied by a
reduction of the phase space which is equivalent to a decrease of
the entropy content of the system. The corresponding states are
stationary and far from equilibrium. Moreover, they are
characterized by dynamic instabilities in the transverse (with
respect to the propagation direction of the GW) microscopic
particle motion. We pointed out the similarity of these properties
to the phenomenon of self-organization.

\ \\
\ \\
{\bf Acknowledgements}\\
\ \\
This paper was supported by the Deutsche Forschungsgemeinschaft
and by the NATO grant
PST. CLG.977973.
\\
\ \\
\ \\

\end{document}